\documentclass[twocolumn,prb,showpacs,superscriptaddress,preprintnumbers,amsmath,amssymb,floatfix,showkeys]{revtex4}

\usepackage[dvips]{graphicx}
\usepackage{xspace}
\usepackage[usenames,dvipsnames]{xcolor}
\usepackage{array}
\usepackage{ulem}
\usepackage{hyperref}
\usepackage[english]{babel}

\newcommand{\COSO}{Cu$_{2}$OSeO$_{3}$\@\xspace}

\newcommand{\ie}{i.\,e.\@\xspace}

\begin{document}

\title{Optically probing symmetry breaking in the chiral magnet Cu$_{2}$OSeO$_{3}$}
\author{R.~B.~Versteeg}
  \affiliation{II. Physikalisches Institut, Universit\"{a}t zu K\"{o}ln, Z\"{u}lpicher Stra{\ss}e 77, D-50937 K\"{o}ln, Germany}
\author{I.~Vergara}
  \affiliation{II. Physikalisches Institut, Universit\"{a}t zu K\"{o}ln, Z\"{u}lpicher Stra{\ss}e 77, D-50937 K\"{o}ln, Germany}
\author{S.~D.~Sch\"{a}fer}
  \affiliation{II. Physikalisches Institut, Universit\"{a}t zu K\"{o}ln, Z\"{u}lpicher Stra{\ss}e 77, D-50937 K\"{o}ln, Germany}
\author{D.~Bischoff}
  \affiliation{II. Physikalisches Institut, Universit\"{a}t zu K\"{o}ln, Z\"{u}lpicher Stra{\ss}e 77, D-50937 K\"{o}ln, Germany}
\author{A.~Aqeel}
  \affiliation{Zernike Institute for Advanced Materials, University of Groningen, Nijenborgh 4, 9747 AG Groningen, The Netherlands}
\author{T.~T.~M. Palstra}
  \affiliation{Zernike Institute for Advanced Materials, University of Groningen, Nijenborgh 4, 9747 AG Groningen, The Netherlands}
\author{M.~Gr\"{u}ninger}
  \affiliation{II. Physikalisches Institut, Universit\"{a}t zu K\"{o}ln, Z\"{u}lpicher Stra{\ss}e 77, D-50937 K\"{o}ln, Germany}
\author{P.~H.~M. van Loosdrecht}
  \affiliation{II. Physikalisches Institut, Universit\"{a}t zu K\"{o}ln, Z\"{u}lpicher Stra{\ss}e 77, D-50937 K\"{o}ln, Germany}
  \email[Corresponding author:]{pvl@ph2.uni-koeln.de}

\date{May 6, 2016}

\begin{abstract}
We report on the linear optical properties of the chiral magnet \COSO, specifically associated with the absence of inversion symmetry, the chiral crystallographic structure, and magnetic order. Through spectroscopic ellipsometry, we observe local crystal-field excitations below the charge-transfer gap. These crystal-field excitations are optically allowed due to the lack of inversion symmetry at the Cu sites.
Optical polarization rotation measurements were used to study the structural chirality and magnetic order. The temperature dependence of the natural optical rotation, originating in the chiral crystal structure, provides evidence for a finite magneto-electric effect in the helimagnetic phase. We find a large magneto-optical susceptibility on the order of 
$\mathcal{V}$($540$\,nm)\,$\sim 10^{4}$\,rad/T$\cdot$m in the helimagnetic phase and a maximum Faraday rotation of $\sim 165$\,deg/mm in the ferrimagnetic phase. The large value of $\mathcal{V}$ can be explained by considering spin cluster formation and the relative ease of domain reorientation in this metamagnetic material. 
The magneto-optical activity
allows us to map the magnetic phase diagram, including the skyrmion lattice phase. 
In addition to this, we probe and discuss the nature of the various magnetic phase transitions in \COSO.
\end{abstract}


\maketitle

\section{Introduction}
Materials with low symmetry exhibit a large variety of intriguing optical phenomena. As is well known, the absence of inversion symmetry leads to the occurrence of natural optical activity. In addition to this, inversion symmetry breaking strongly affects the electric dipole selection rule, and thus allows for direct excitation of local crystal-field excitations in linear spectroscopy.
In magnetic materials, the breaking of time-reversal symmetry leads to the magneto-optical Kerr and Faraday effect. Many of the optical phenomena originating from
a reduced symmetry have technological applications, examples of
which include optical insulators and
magneto-optical storage devices. \cite{zvezdin}

In recent years, materials lacking both spatial inversion and time-reversal symmetry have been a focal point of condensed matter research. Their combined absence may for instance lead to magneto-electric and multiferroic behavior,\cite{multiferroicreviewtokura2014, khomskii2009, Khomskii2006} the formation of chiral and skyrmion magnetic ground states,\cite{tokura2013} and the occurrence of toroidal order, \cite{spaldin2008,zimmermann2014ferroic} and excitations. \cite{Kaelberer2010,papasimakis2016toroidal} Effects on the optical properties have been studied extensively.\cite{krichevtsov1993,pimenov2006possible,rovillain2012} One intriguing optical phenomenon originating from the combined absence of time-reversal and space-inversion symmetry is the so-called nonreciprocal directional dichroism.\cite{szaller2013,kezsmarki2014,toyoda2015}

The high sensitivity of optical properties to symmetry breaking may be used to gain a better understanding of the underlying material properties. Here we focus on the cuprate material \COSO, belonging to the class of non-centrosymmetric cubic crystal structures. These materials have recently triggered a great deal of research interest owing to the occurrence of topologically protected spin-vortex-like structures, known as skyrmions.\cite{tokura2013}

In these chiral crystal structures, the absence of inversion symmetry between the spin sites leads to a non-vanishing antisymmetric Dzyaloshinskii-Moriya (DM) exchange interaction, which competes with the energetically stronger isotropic Heisenberg exchange. This combination of exchange interactions stabilizes spin helices.\cite{bak1980} In the presence of an external magnetic field, the Zeeman interaction energy stabilizes the formation of a topologically robust hexagonal lattice of nanometer-sized skyrmions.\cite{roessler2011}

In this article we report on a variety of linear optical properties \cite{glazer} associated with the broken inversion symmetry, structural chirality, and magnetic order in the chiral magnet \COSO.\cite{seki2012} The orbital aspect of \COSO has only received minimal experimental attention so far.\cite{langner2014} Here, we reveal local crystal-field excitations below the charge-transfer gap by means of spectroscopic ellipsometry. These orbital excitations acquire a finite dipole matrix element due to the low crystal-field symmetry. Optical polarization rotation measurements are used to study the structural chirality and to probe the magnetic order. The natural optical activity, resulting from the chiral crystal structure, shows an abrupt change upon magnetic ordering. This observation is evidence of a finite magneto-electric coupling in the phase with helical magnetic order.
The large magneto-optical response is quantified by a magneto-optical susceptibility on the order of $\mathcal{V}$($540\,$nm)\,$\sim$\,10$^{4}$\,rad/T$\cdot$m in the helimagnetic phase, and maximum Faraday rotation of $\sim 165\,$deg/mm in the field-polarized ferrimagnetic phase. This 
strong
response serves as an excellent probe for the various magnetic phase transitions. The magneto-optical data allow us to derive the phase diagram of \COSO as a function of magnetic field and temperature. The optically determined phase diagram is in excellent agreement with results obtained by other techniques.\cite{longwavelength,seki2012,ruff2015}

\section{Structure and magnetism}

\begin{figure}[t]
 \center
\includegraphics[width=3.375in]{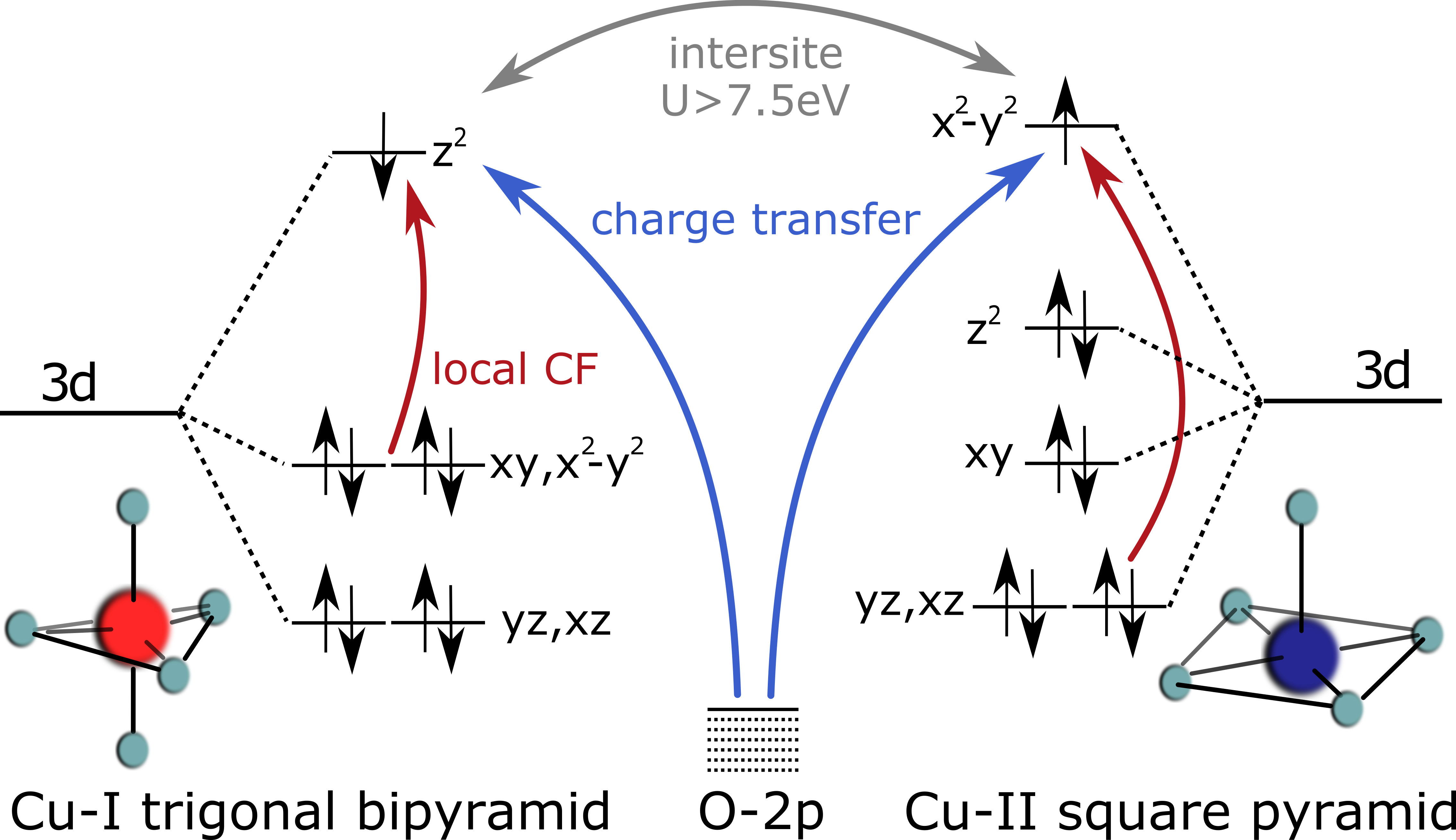} 
\caption{The $3d$ crystal-field splittings for the Cu-I (trigonal bipyramid) and Cu-II (square pyramid) ions. The dipole-active local crystal-field (CF) excitations are indicated with red arrows, and the charge-transfer excitations with blue arrows. Intersite excitations between the Cu ions, illustrated with a gray arrow, lie outside the experimentally accessible energy range of this work.}
\label{fig:splittings}
\end{figure}

\COSO has a complex chiral crystal structure with cubic space group $P2_{1}3$. The unit cell contains 16 Cu ions, all having a $2+$ valence state ($d^9$ configuration). These Cu ions are located within two crystallographically distinct oxygen ligand field geometries,\cite{jwbos2008} which can be approximated by a trigonal bipyramid ($D_{3h}$) for Cu-I ions and by a square pyramid ($C_{4v}$) for Cu-II ions, as depicted in Fig.\ \ref{fig:splittings}. The true site symmetries are lower, and are given by $C_{3v}$ for Cu-I and $C_{1}$ for the Cu-II ions.\cite{jwbos2008} These ligand fields lead to different crystal-field splittings of the $3d$ orbitals on the Cu-I and Cu-II sites, as shown in Fig.\ \ref{fig:splittings}. For Cu-I the hole is located in the ${z^2}$ orbital, whereas for Cu-II the hole is in the ${x^2-y^2}$ orbital. Note that these are no pure $3d$ orbitals due to the low site symmetry, \ie, parity is not a good quantum number. 

The 16 Cu ions in the unit cell are distributed over 4 Cu-I sites and 12 Cu-II sites and form a network of corner-sharing tetrahedra, where each Cu tetrahedron comprises 1 Cu-I and 3 Cu-II ions. Owing to the different bond lengths between the 16 Cu$^{2+}$ sites in the unit cell, it is possible to make a real space classification of the exchange energy scales into a "strong" and "weak" type.\cite{clustercoso} Through the strong Heisenberg and DM exchange interactions, tetrahedral 3-spin-up-1-down triplet clusters couple far above\cite{clustercoso} the macroscopic ordering temperature $T_C$\,$\approx$\,$58$\,K. These $S$\,=\,$1$ spin clusters turn out to be the relevant low-energy spin entities in \COSO. The weaker inter-tetrahedral Heisenberg and DM exchange 
couplings mediate the interactions between the $S$\,=\,$1$ entities, giving rise to long-range helimagnetic order below $T_C$. This separation into inter- and intra-tetrahedral energy scales is well supported by the splitting of the magnon spectra in two well separated energy bands.\cite{portnichenko2016}
Finally, weak cubic magnetic anisotropy terms pin the helimagnetic spirals along the six equivalent
crystallographic $\langle$100$\rangle$ directions, leading to domain formation.\cite{longwavelength,belesi2012,magnetoelectricnature}
In the presence of an external magnetic field, different metamagnetic phases are formed. Applied magnetic fields of a few tenths of mT are enough to fully lift the degeneracy of the helical domains, giving a conical type of order with the propagation vector $q$ along the applied field. Above a second critical field the system is driven into the field-polarized ferrimagnetic phase, where all tetrahedral $S$\,=\,$1$ entities are aligned with the magnetic field. The skyrmion lattice phase (SkL) is located within a narrow field-temperature window just below $T_C$\,$\approx$\,$58$\,K for moderate applied fields of the order of 20\,-\,50\,mT.\cite{longwavelength,seki2012}

\section{Experimental methods}

High-quality single crystals were grown using a standard chemical vapor transport method.\cite{belesi2010ferrimagnetism} The studied crystals were oriented by crystal morphology inspection after which the fine orientation was done by means of a Laue camera. For the ellipsometry study a (100) surface was prepared (sample surface dimensions approximately 2.7\,mm\,$\times$\,2.8\,mm), whereas for the polarimetry study a (111) oriented sample was chosen (surface area 3.6\,mm\,$\times$\,3.2\,mm, used thicknesses $d$\,$\approx$\,1.00\,mm and $d$\,$\approx$\,221\,$\pm$\,3\,$\mu$m).
The samples were polished with Al$_2$O$_3$ suspension ($\approx 1\,\mu$m grain size) in order to obtain optically smooth surfaces.

For the optical spectroscopy part, a Woollam VASE spectroscopic ellipsometer with an autoretarder between source and sample was used. Ellipsometry allows us to obtain the real and imaginary parts of the complex dielectric function in a self normalizing way. 
The (100) sample was mounted in a UHV chamber with liquid-He flow cryostat and measured in the range $0.75$ to $5$\,eV at a fixed angle of incidence ($70^\circ$). For the analysis of the ellipsometric data, the surface roughness was estimated using the knowledge that absorption is small in the transparency windows below 1\,eV and around $2.3\,$eV.\@

For the polarization rotation measurements we used a homebuilt optical polarimetry setup based on the polarization modulation technique described in Refs.\ \onlinecite{satopem} and \onlinecite{polisetty2008}. Measurements are possible in the energy range of $1.1$\,eV to $3.5$\,eV in fields up to $5$\,T and temperatures down to $10$\,K.\@
The measurements are performed in Faraday (transmission) geometry, where the light propagates along the crystallographic [111] direction, with the magnetic field also applied along this direction.

\raggedbottom

\section{Zero-field optical properties}

\subsection{Optical excitations}

\begin{figure}[t]
\center
\includegraphics[scale=1]{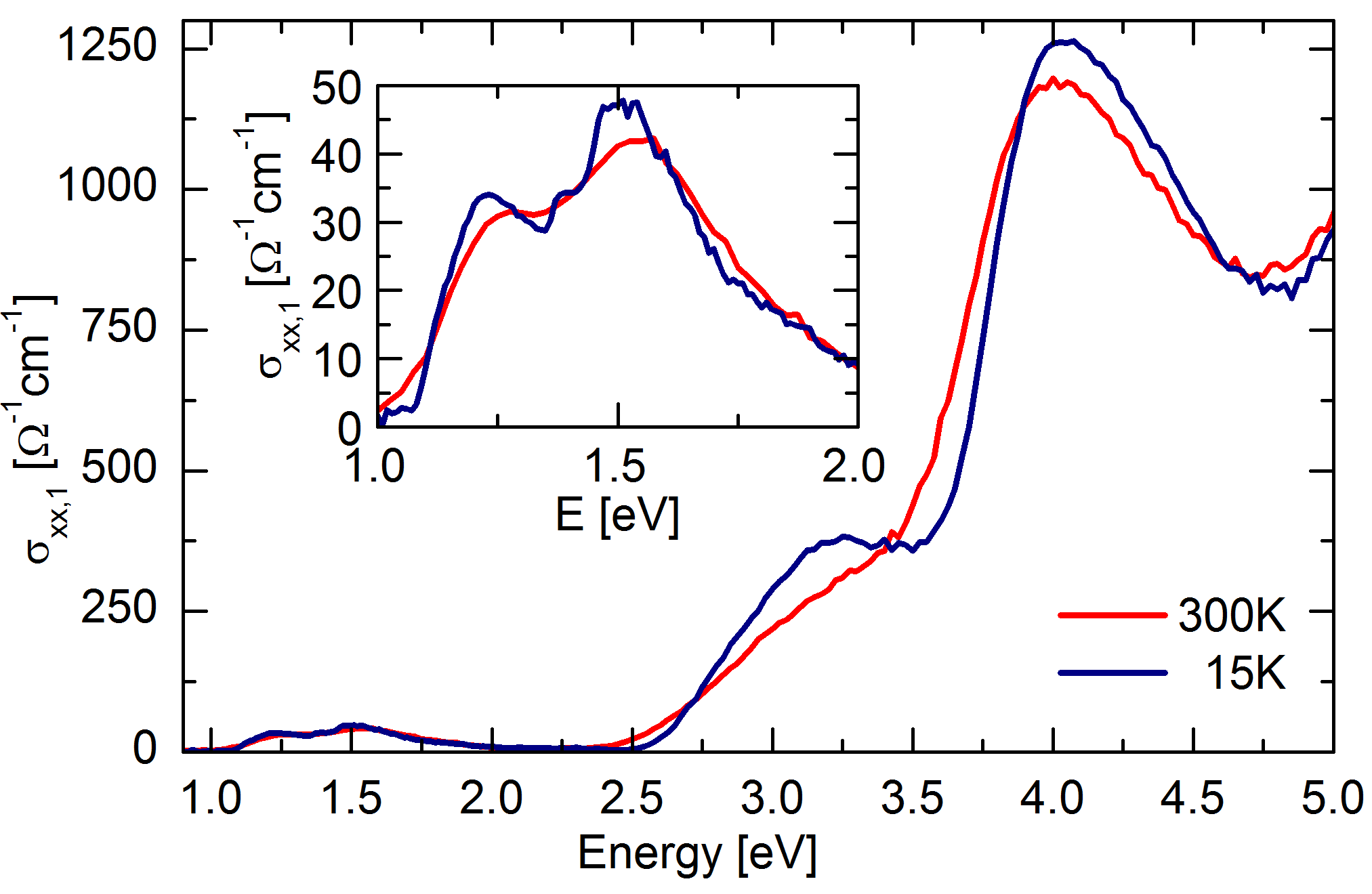}
\caption{Optical conductivity at $15$\,K and $300$\,K as obtained by ellipsometry. The onset of charge-transfer excitations is observed at about $2.5$\,eV at $15$\,K.\@ 
Between $1.0$ and $2.0$\,eV local crystal-field excitations are observed with conductivity values not surpassing $50$\,$\Omega^{-1}$cm$^{-1}$. At $15$\,K, pronounced peaks are located at $1.2$, $1.4$, and $1.5$\,eV.\@ The features broaden with increasing temperature, obscuring the $1.4$\,eV peak at $300$\,K, as seen in the inset.}
\label{fig:COSO-ellip-full}
\end{figure}

In Fig.\ \ref{fig:COSO-ellip-full} the diagonal component $\sigma_{xx,1}$ of the optical conductivity at $15$\,K and $300$\,K is shown. We find a clear electronic gap with an onset at about $2.5$\,eV at $15$\,K, and charge-transfer excitations peaking at $3.2$\,eV and $4.0$\,eV with an optical conductivity of about $400$ and $1200$\,$\Omega^{-1}$cm$^{-1}$ respectively. DFT+$U$ calculations find narrow Cu hole bands for both Cu sites with an energy difference of about $\Delta E$\,=\,$E_{II}-E_{I}$\,=\,0.2\,eV, while the valence band primarily consists of broad oxygen bands.\cite{yang2012} Based on this, we tentatively
attribute the structure in the charge-transfer region to both, the splitting $\Delta E$ and the structure in the O $2p$ valence band density of states. 
With decreasing temperature, the charge-transfer excitations show a blueshift and a sharpening.

The first consequence of the low structural symmetry appears in the optical conductivity. Between $1.0$ and $2.0$\,eV we find a multi-peak absorption feature, with a conductivity maximum of about $50$\,$\Omega^{-1}$cm$^{-1}$, \ie, much weaker than the charge-transfer excitations (see inset of Fig.\ \ref{fig:COSO-ellip-full}).  At 15\,K, pronounced peaks are located at $1.2$, $1.4$, and $1.5$\,eV.\@ The peaks broaden with increasing temperature, obscuring the peak at $1.4$\,eV at $300$\,K.\@ We interpret these features as local crystal-field (CF) excitations. 
Similar excitation energies were observed before for CF excitations of Cu$^{2+}$ ions in a trigonal bipyramidal crystal field.\cite{Kuroda81}
For materials with inversion symmetry at the transition-metal site, such  crystal-field excitations would be parity forbidden within the dipole approximation. They only become weakly allowed via the simultaneous excitation of an inversion-symmetry-breaking odd-parity phonon. Typical values of $\sigma_1(\omega)$ for such phonon-assisted excitations are below $10$\,$\Omega^{-1}$cm$^{-1}$.\cite{ruckkamp} In contrast, the absence of inversion symmetry at the Cu-I and Cu-II sites in \COSO allows for dipole-active CF excitations by symmetry, and thus naturally explains the 
sizable spectral weight below the gap between $1.0$ and $2.0$\,eV.

According to group theory,\cite{POINT} dipole-active excitations are allowed from $xy$ and $x^{2}-y^{2}$ to $z^{2}$ for the Cu-I site (assuming $D_{3h}$ symmetry) and from $xz$ and $yz$ to $x^{2}-y^{2}$ for the Cu-II site (assuming $C_{4v}$ symmetry). 
Both sets of transitions are schematically indicated in Fig.\ \ref{fig:splittings}.
The spectral weight in the low-energy region, however, has a richer structure with at least three peaks at $15$\,K.\@ In the above mentioned analysis we ignored the fact that the Cu sites show a slight distortion away from the ideal square pyramidal and trigonal bipyramidal 
symmetries. Considering the correct, lower site symmetries, the
remaining crystal-field excitations also are dipole active and will also contribute to the total spectral weight. Note that the two peak energies observed at 300\,K \textit{soften} to lower energy with decreasing temperature 
(see the appendix for a more detailed $T$ dependence).
This is opposite to the behavior of the charge-transfer peaks and can be attributed to the asymmetric line shape found in particular at low temperature. Such an asymmetric line shape can be described by the Franck-Condon line shape typical for crystal-field excitations.

\begin{figure}[t] 
\center
\includegraphics[scale=1]{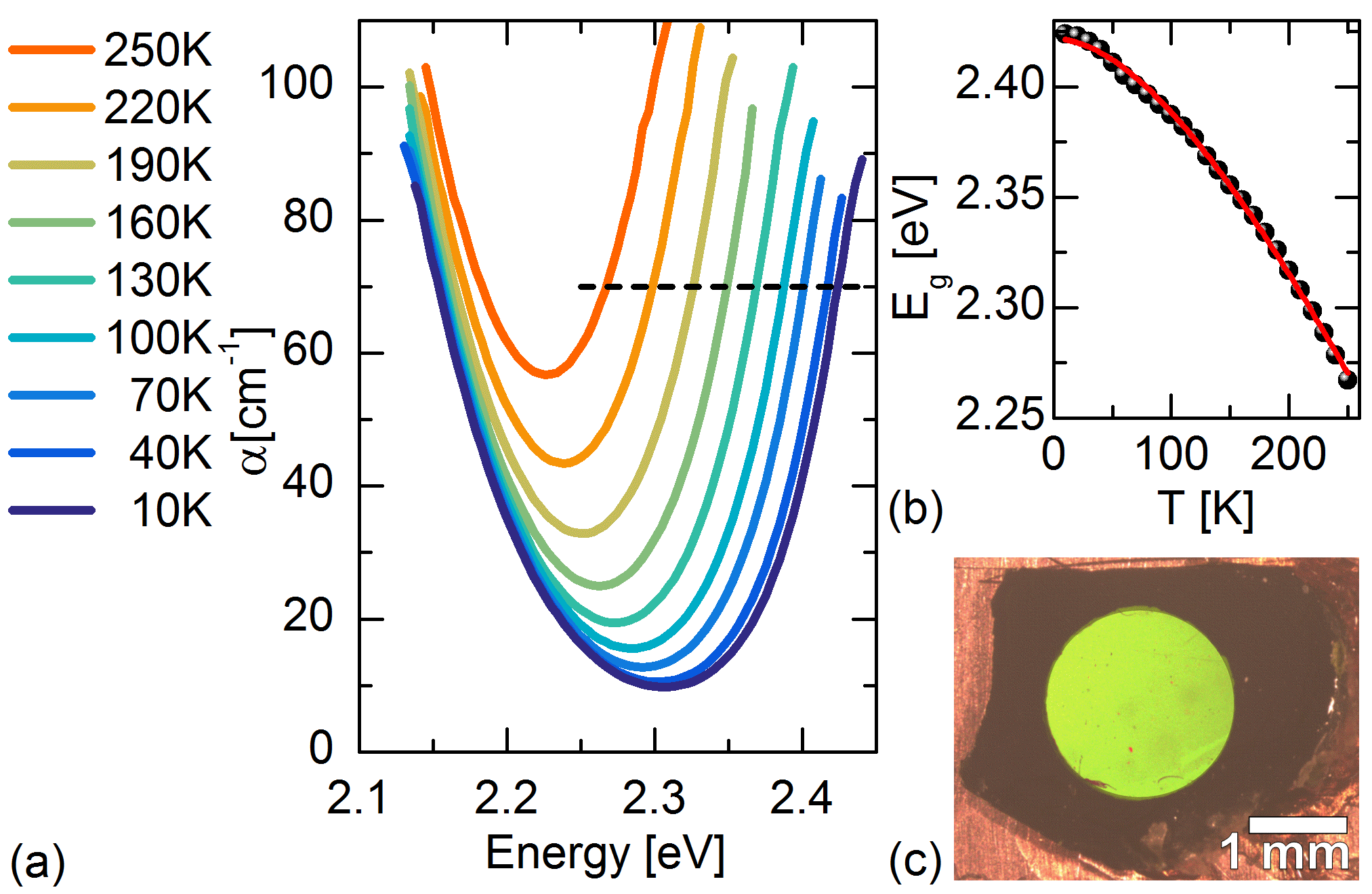}
\caption{(a) Temperature-dependent absorption spectra in the range of the transmission window as measured on the $1$\,mm thick sample.
(b) Temperature dependence of the onset $E_g$ of excitations across the gap, determined at $\alpha$\,=\,$70$\,cm$^{-1}$. The fit (red line) is based on the empirical Varshni equation. (c) \COSO is transparent for green light. The (111) sample, with a thickness $d$\,$\approx$\,$221$\,$\pm3\,\mu$m, is glued on a copper sample holder with a 2.0\,mm diameter hole.}
\label{fig:absorptioncoeff}
\end{figure}

Our assignment of the onset of charge-transfer excitations to an energy of about $2.5$\,eV is corroborated by the observation of a narrow transmission window which is situated between the crystal-field excitations and the charge-transfer region. At $10$\,K the absorption coefficient becomes as low as $10$\,cm$^{-1}$ at $2.3$\,eV, see Fig.\ \ref{fig:absorptioncoeff}. On the low-energy side the absorption decreases with decreasing temperature due to the sharpening of the crystal-field excitations. On the high-energy side a more drastic change  of the absorption coefficient is found, reflecting the temperature dependence of the onset $E_g$ of excitations across the gap. This temperature dependence is well described by the empirical Varshni equation,\cite{varshni1967} as seen in Fig.\ \ref{fig:absorptioncoeff}b.

Previously, Miller \textit{et al}.\cite{Miller2010} reported on the optical conductivity of \COSO at 300\,K based on a Kramers-Kronig analysis of reflectivity data. Above 2.5\,eV, their data reasonably agree with our results, showing a dominant peak at 4\,eV.\@ However, a Kramers-Kronig analysis is not sensitive to weak absorption features, thus the crystal-field excitations between $1.0$ and $2.0$\,eV were not resolved. 
Additionally, Miller \textit{et al}.\cite{Miller2010} reported on a transmission window in the frequency range above the phonons and below about 1\,eV, which agrees with the onset of crystal-field excitations observed in our data. 

\subsection{Natural optical activity}

\begin{figure}[t]
	\center
	\includegraphics[scale=1]{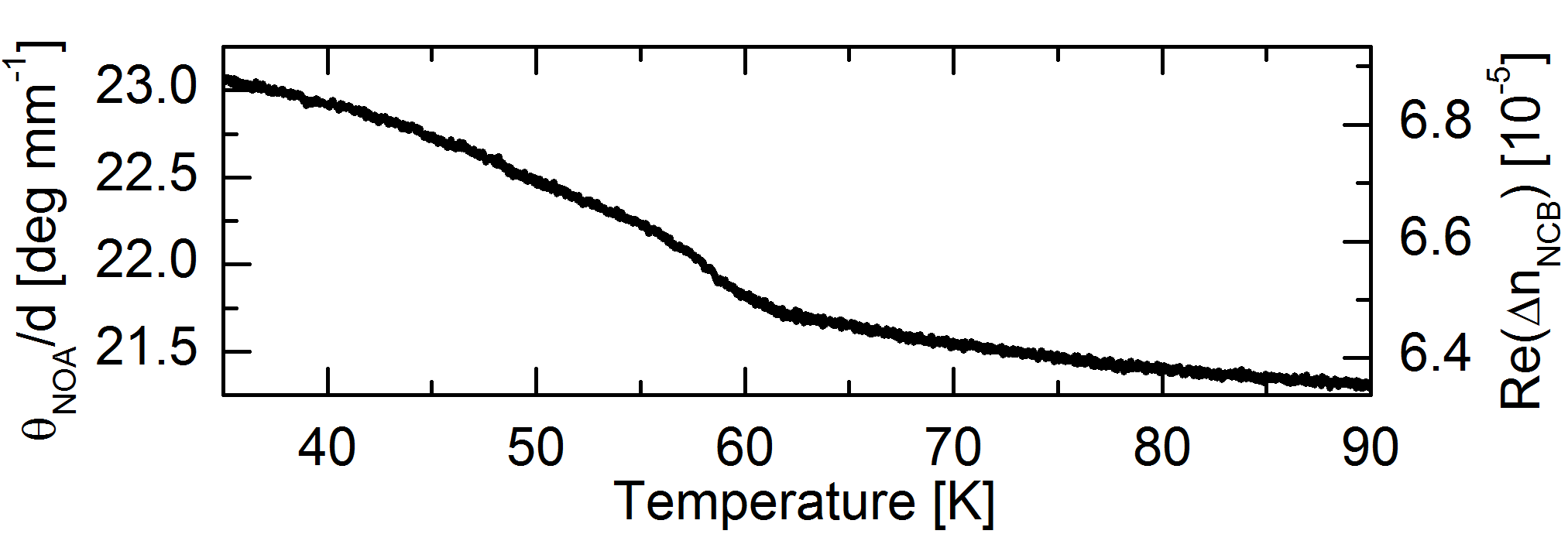} 
	\caption{The natural optical activity measured along the crystallographic (111) axis with probe photon energy $E_{\rm probe}$\,=\,$2.30\pm0.03$\,eV.\@ The left axis gives the natural optical rotation $\theta_{\rm NOA}$ per sample thickness $d$ in deg/mm and the right axis the natural circular birefringence (NCB). Above the Curie temperature $T_c\approx58$\,K, the temperature dependence is weak. Below $T_c$ an enhancement of $\theta_{\rm NOA}$ is observed, which is attributed to the  finite magneto-electric coupling in the helimagnetically ordered state.}
	\label{fig:temperaturencb}
\end{figure}

Due to the chiral crystal structure, \COSO is expected to show circular birefringence with a concomitant optical rotation for linearly polarized light, known as natural optical activity with rotation angle $\theta_{\rm NOA}$.\cite{glazer,BARRONbook2004} The low absorption around $2.3$\,eV allows us to measure the temperature dependence of $\theta_{\rm NOA}$ across the paramagnetic-helimagnetic phase transition ($T_C$\,$\approx$\,$58$\,K) in transmission geometry. As a probe wavelength we used $\lambda_{\rm probe}$\,=\,$540\pm5$\,nm ($E_{\rm probe}$\,=\,$2.30\pm0.03$\,eV), which corresponds to the low-temperature transmission maximum.
The result is given in Fig.\  \ref{fig:temperaturencb}, showing both $\theta_{\rm NOA}/d$
and the difference of the refractive indices for left and right circularly polarized light, Re($\Delta n_{\rm NCB}$), where NCB is natural circular birefringence. Above  $T_C$ a finite $\theta_{\rm NOA}/d$ of around $21.5$\,deg/mm is found. The rotation sign shows that the studied (111) oriented \COSO sample is dextrorotatory, as seen from the source.\cite{newnham}
The temperature dependence of $\theta_{\rm NOA}$ in the paramagnetic phase hints at an increase of the structural chirality upon lowering temperature, \ie, a displacement of ions at general coordinates within the \COSO unit cell, satisfying the threefold rotational symmetry of the structural helix.

In the helimagnetically ordered phase, the temperature dependence of $\theta_{\rm NOA}$ is enhanced. Since there is no net magnetization in the helimagnetic 
phase for $B$\,=\,0,\cite{belesi2012} no Faraday rotation is expected. Different studies suggest that no significant magnetostrictive lattice contraction nor lattice deformations are present in \COSO related to the transition into the 
helimagnetic phase.\cite{kurnosov2012, jwbos2008} 
Instead, the increasing $\theta_{\rm NOA}$ strongly hints to the presence of a finite magneto-electric coupling in the helimagnetic phase, as suggested in Refs. \onlinecite{ruff2015}, \onlinecite{jwbos2008}, \onlinecite{belesi2012} and \onlinecite{magnetoelectricnature}. 

\section{Magneto-optical properties}

\subsection{Phase transitions}

In the presence of a magnetic field $B_a$\,$\parallel$\,[111], different meta\-magnetic phases with a finite magnetization $M(B_a)$ form in \COSO. For these phases, the total polarization rotation is of the form
\begin{equation}
\theta_{\rm tot} = \theta_{\rm NOA} + \theta_{\rm F}(M(B_a)) + \theta_{\rm HO}(M^2(B_a^2))  \,\,\, . 
\label{eq:thetatot}
\end{equation}
The last term gives a higher-order (HO) rotation in field which is discussed later in the text, 
while $\theta_{\rm F}$ denotes the Faraday rotation, which is proportional to the magnetization 
$M(B_a)$ in the [111] direction.\cite{pershan1967} 
The field-dependent magnetization can be rewritten as $M(B_a)$\,=\,$\chi_m(B_a)$\,$\cdot$\,$B_a$, where $\chi_m(B_a)$ refers to the magnetic susceptibility. Note that the magnetic susceptibility $\chi_m$ itself is 
a function of the external field. We thus probe a Faraday rotation per sample thickness $d$ of
\begin{equation}
\begin{split}
\frac{1}{d}\,\theta_{\rm F}(\omega,M(B_a)) & = \beta(\omega)\cdot M(B_a)\\  
& = \beta(\omega)\cdot\chi_m(B_a)\cdot B_a \\
\end{split}
\label{eq:faradayrotation}
\end{equation}
where $\beta(\omega)$ captures the microscopic magneto-optical properties. The Faraday rotation is anti-symmetric in applied field. By making field sweeps between opposite polarity field values, and afterwards symmetrizing/anti-symmetrizing the rotation response, $\theta_{\rm F}(M(B_a))$ is obtained. The symmetric part gives $\theta_{\rm NOA}+\theta_{\rm HO}(M^2(B_a^2))$. The used probe wavelength is again $\lambda_{\rm probe}$\,=\,$540\pm5$\,nm. 

\begin{figure}[t]
\center
\includegraphics[scale=1]{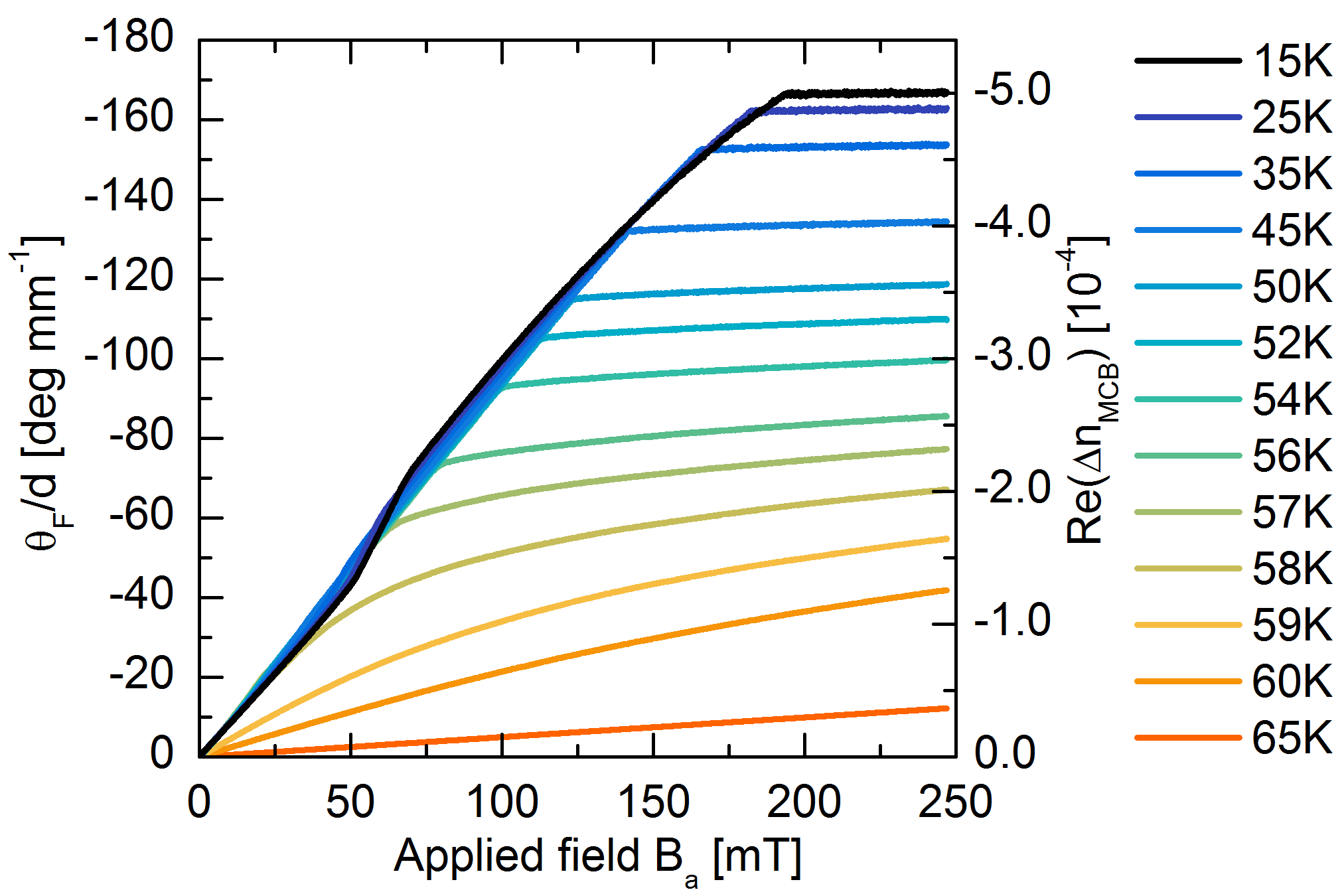} 
\caption{The Faraday rotation per sample thickness, $\theta_{\rm F}/d$, as a function of field for different temperatures. The right axis gives the corresponding magnetic circular birefringence (MCB). With increasing field, $\theta_{\rm F}$ increases until a plateau is reached, marking the 
phase transition between conical and ferrimagnetic order.
At $15$\,K this phase transition is induced around an applied field of $B_a$\,$\approx$\,195\,mT.\@ At lower fields, in the $15$\,K curve around  $B_a$\,$\approx$ 60\,mT, the helical-conical phase transition becomes apparent as a kink in the field dependence 
of $\theta_{\rm F}$.}
\label{fig:faradayrotation}
\end{figure}

Figure \ref{fig:faradayrotation} shows the Faraday rotation per sample thickness, $\theta_{\rm F}/d$, as a function of field $B_a$ for temperatures ranging from $15$\,K to $65$\,K.\@ At zero applied field, different helimagnetic domains exist. However, there is no net magnetization, even for a single helical domain. In this way, $\theta_{\rm F}$ is zero for $B_a$\,=\,$0$\,mT.\@ With increasing field the helimagnetic domains acquire a conical contribution, leading to a finite field-induced magnetization $M(B_a)$, and hence a Faraday rotation. The multi-$q$ helimagnetic domain structure however still persists. At a critical field of around $60$\,mT at $15$\,K, the reorientation transition from the multi-$q$ helical to the single-$q$ conical phase is observed. 
With increasing field, the spin projection parallel to $q$ and $B_a$ increases. It is for this reason that in the conical phase the Faraday rotation still increases with field, until a plateau is reached, marking the phase transition from conical to field-polarized ferrimagnetic order. At $15$\,K this phase transition is induced 
around an applied field of $B_a$\,$\approx$\,$195$\,mT.\@ At the plateau a rotation of around $165$\,deg/mm is found for $15$\,K. The Faraday rotation sense 
was determined to be 
levorotatory as seen from the source.

\begin{figure}[t]
\centering
 \includegraphics[width=3.375in]{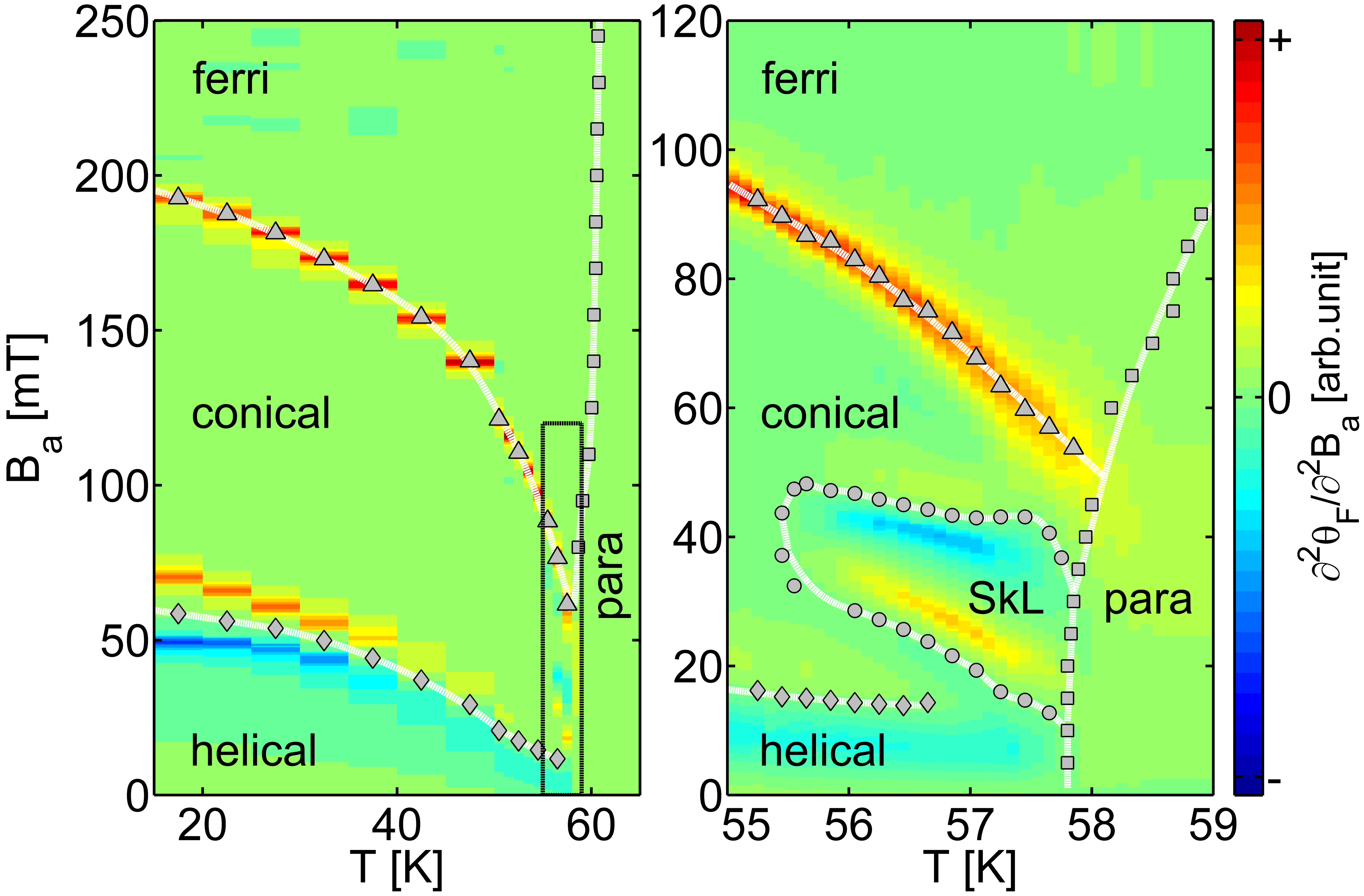}
 \caption{The magnetic phase diagram of \COSO as obtained from the Faraday rotation. $B_a$ is the applied magnetic field  along the crystallographic (111) direction. The left panel shows the magnetic phase diagram for a large ($B_a$,$T$) range, whereas the right panel gives a zoom-in around the skyrmion lattice phase (SkL). The triangles indicate the conical-ferrimagnetic phase boundary. Diamonds give the helical-conical phase boundary. The SkL phase boundary is indicated by the circles. The phase boundary between ordered phases and the paramagnetic phase is indicated by the squares. The color mapping indicates the second derivative of the Faraday rotation. 
For a quantitative comparison of the critical fields with results obtained by other techniques, 
one needs to take into account the demagnetization factor (see main text).
 }
\label{fig:phasediaglarge}
\end{figure}

The second derivative\cite{numdiffchartrand} of the Faraday rotation allows us to construct the phase diagram shown in Fig.\ \ref{fig:phasediaglarge}. The right panel gives a zoom-in around $T_C$. In this way the phase transitions to the skyrmion lattice phase (SkL) become apparent. With a field applied along the $\langle$111$\rangle$ hard axis, the phase transition from conical to ferrimagnetic order (indicated by triangles) is of second order,\cite{clustercoso} 
and the phase transition can be identified with
a local maximum in the second derivative of the order parameter $M$\,$\propto$\,$\theta_{\rm F}$. The phase transitions from helical to conical order (diamonds) and between the conical phase and the SkL phase (indicated by circles) are of first order 
and can be identified with
a zero-crossing in the second derivative of $\theta_{\rm F}$. The diamonds and circles in Fig.\ \ref{fig:phasediaglarge} indicate these zero crossings, marking the phase boundaries.

\begin{figure}[htb]
\center
\includegraphics[scale=1]{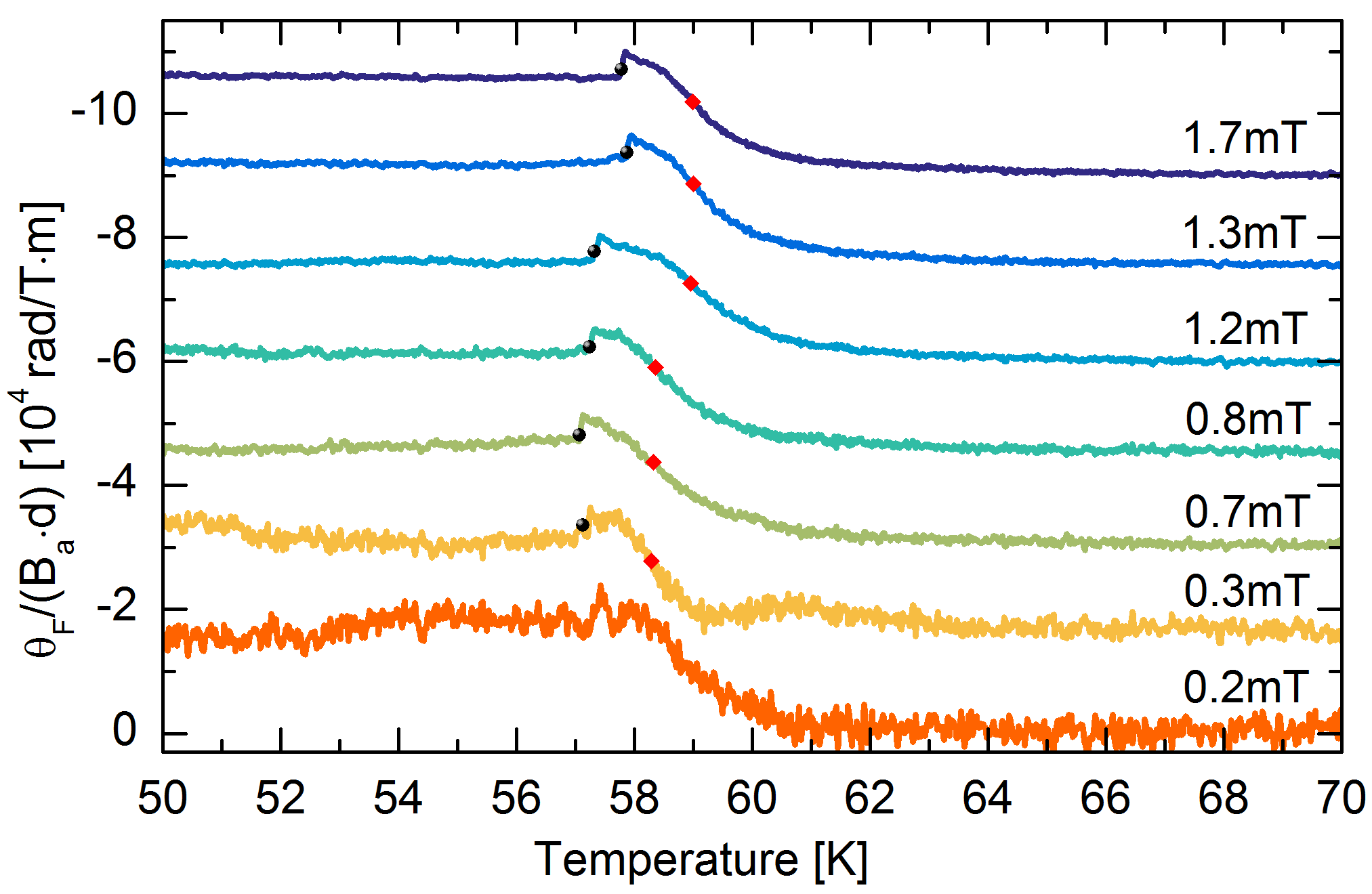} 
\caption{Faraday rotation across the paramagnetic-helimagnetic phase transition, normalized by applied field $B_a$. 
The curves are offset by $-(\pi/1.8)\cdot 10^5$\,rad/T$\cdot$m with respect to each other. Within the helical phase the scaled rotation attains a value of $\theta_{\rm F}$\,/($B_a$\,$\cdot$\,$d$)\,$\approx$\,$-1.7\cdot10^5$\,rad/T$\cdot$m.
The weak first-order transition into the helical phase is marked by black circles. The fluctuation-disorder regime is located between the black dot and the red diamond.
}
\label{fig:fluctuationtemperature}
\end{figure}

The phase transitions from the paramagnetic phase to the ordered phases are best observed in temperature sweeps at constant field. Depending on the field strength, different phase transitions occur. 
Figure \ref{fig:fluctuationtemperature} shows $\theta_{\rm F}$\,/($B_a$\,$\cdot$\,$d$) across the paramagnetic-helimagnetic phase transition for $\vert B_a\vert$\,\textless\,$2$\,mT.\@ At low temperatures, deep within the helimagnetic phase, $\theta_{\rm F}$\,/($B_a$\,$\cdot$\,$d$) does not show a significant temperature dependence. However, 
an anomaly
is observed around the paramagnetic-helimagnetic transition. In fact, the 
anomaly
marks $T_C$. In mean-field approximation this phase transition is expected to be of second order. However, the interaction between chiral fluctuations renormalizes the phase transition into a weak first-order transition,\cite{firstordergarst} as seen in the temperature dependence of $\theta_{\rm F}$\,/($B_a$\,$\cdot$\,$d$). 
Just above the phase transition, we find the fluctuation-disordered region \cite{Bauer2016} where the magnetic susceptibility deviates from pure Curie-Weiss behavior. The onset of this region is identified by the inflection point, located about $1$\,K above $T_C$ (indicated by the red diamonds).  

\begin{figure}[t]
\center
\includegraphics[scale=1]{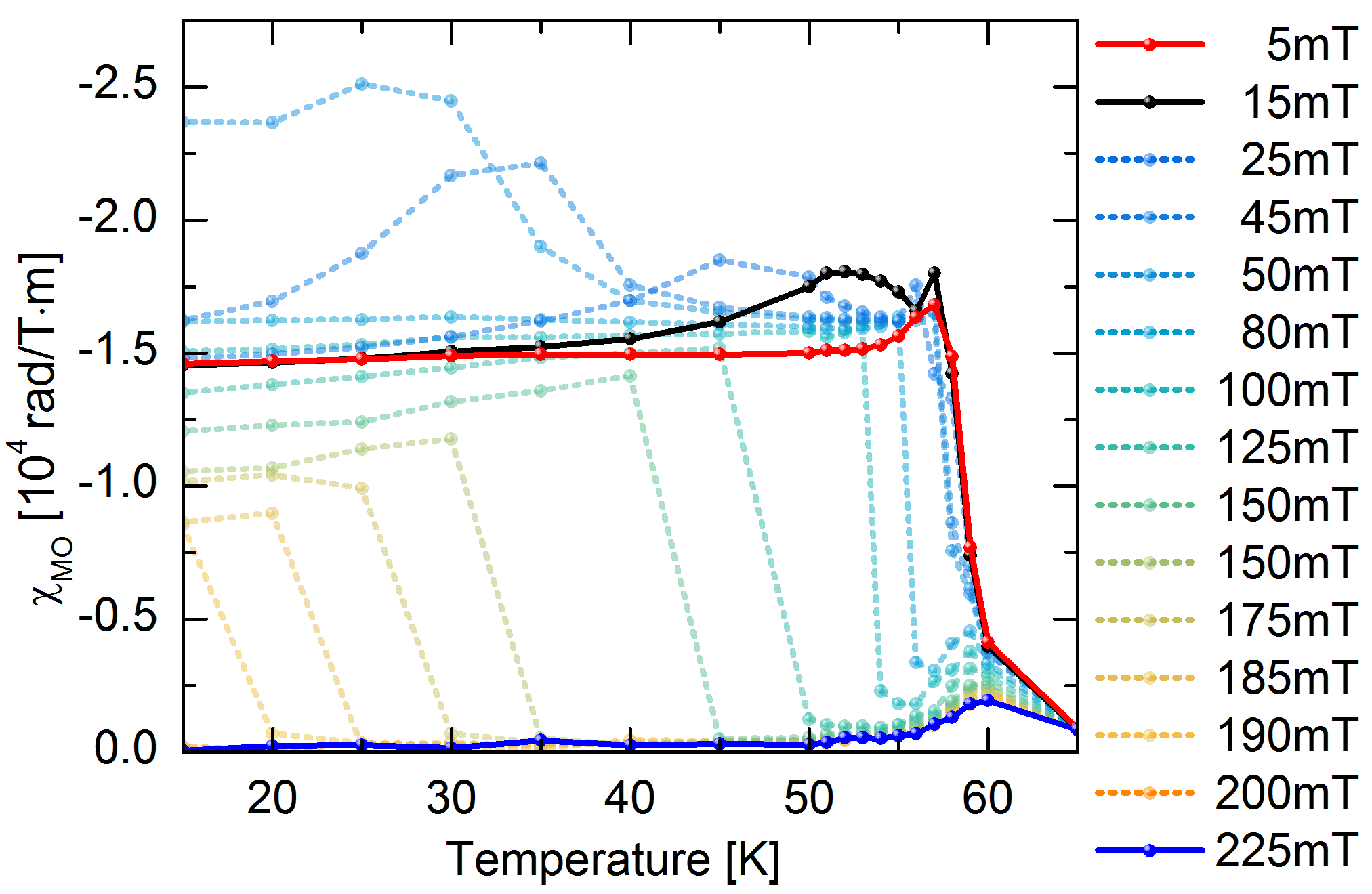} 
\caption{The magneto-optical susceptibility $\chi_{\rm MO}$ as a function of temperature at different applied field strengths. The phase transition from the paramagnetic phase to 
one of
the ordered phases shows a rescaling from first 
order
(for the para to helical/conical/SkL phases) to second order (para-ferri) with increasing field. The changes in $\chi_{\rm MO}$ below $T_C$\,$\approx$\,$58$\,K show the temperature-induced conical-helical and ferrimagnetic-conical phase transitions.}
\label{fig:fluctuationsverdet}
\end{figure}

The high-field phase transitions from the paramagnetic to the ordered phases are best seen in the temperature dependence of the first derivative $\partial \theta_{\rm F}/\partial B_a$, at different values of $B_a$. 
Figure \ref{fig:fluctuationsverdet} shows
\begin{equation}
\chi_{\rm MO}=\frac{1}{d}\frac{\partial \theta_{\rm F}}{\partial B_a}\,=\,\beta(\omega)\big[\chi_m(B_a)+ \frac{\partial \chi_m}{\partial B_a}B_a\big]
\label{eq:susceptibility}
\end{equation}
given in rad/T$\cdot$m. 
In this way the first derivative 
can be identified with the
magneto-optical (MO) susceptibility.
The temperature and field dependence 
of $\chi_{\rm MO}$
show good agreement with the magnetic ac-susceptibility as reported by \v{Z}ivkovi\'{c} \textit{et al}. \cite{criticalscaling} 

For $5$\,mT the characteristic 
anomaly
for the para\-magnetic-helimagnetic transition is visible again. For applied fields of $15$, $25$, and $45$\,mT the first-order nature remains for the paramagnetic-SkL and paramagnetic-conical transitions. The broad maximum found at lower temperatures for these applied fields indicates the temperature-induced 
conical-helical
phase transition. At higher fields such as $100$, $150$, and $225$\,mT, the paramagnetic-ferrimagnetic phase transition is crossed. The magneto-optical susceptibility for these field values shows that the character of the phase transition changes to second order. For \COSO, this 
change of the phase-transition type
has been discussed extensively by \v{Z}ivkovi\'{c} \textit{et al}.\cite{criticalscaling} 
The increase in $\chi_{\rm MO}$ at lower temperatures for field values between $80$ and $190$\,mT shows 
where the temperature-induced ferrimagnetic-conical phase transition is crossed. With the first derivative, 
the phase diagram can be completed. The boundary between the paramagnetic phase and the ordered phases
is indicated by squares in the phase diagram. 

Qualitatively, the phase diagram that we have determined by optical means is in excellent agreement with previously reported results based on other techniques.\cite{seki2012,longwavelength,ruff2015} For a quantitative comparison of the applied field strengths at which the phase transitions are observed, one has to take into account demagnetization effects related to the sample shape. For the optical measurements in transmission geometry, we used a thin plate for which demagnetization effects are substantial. The demagnetization factor can be estimated \cite{satoishii1989} to lie around $N_z$\,$\approx$\,$0.9$. Accordingly, the phase transitions occur at higher applied fields than for instance in a spherical sample with $N_z$\,$=$\,$1/3$. 

\begin{figure}[t]
\center
\includegraphics[scale=1]{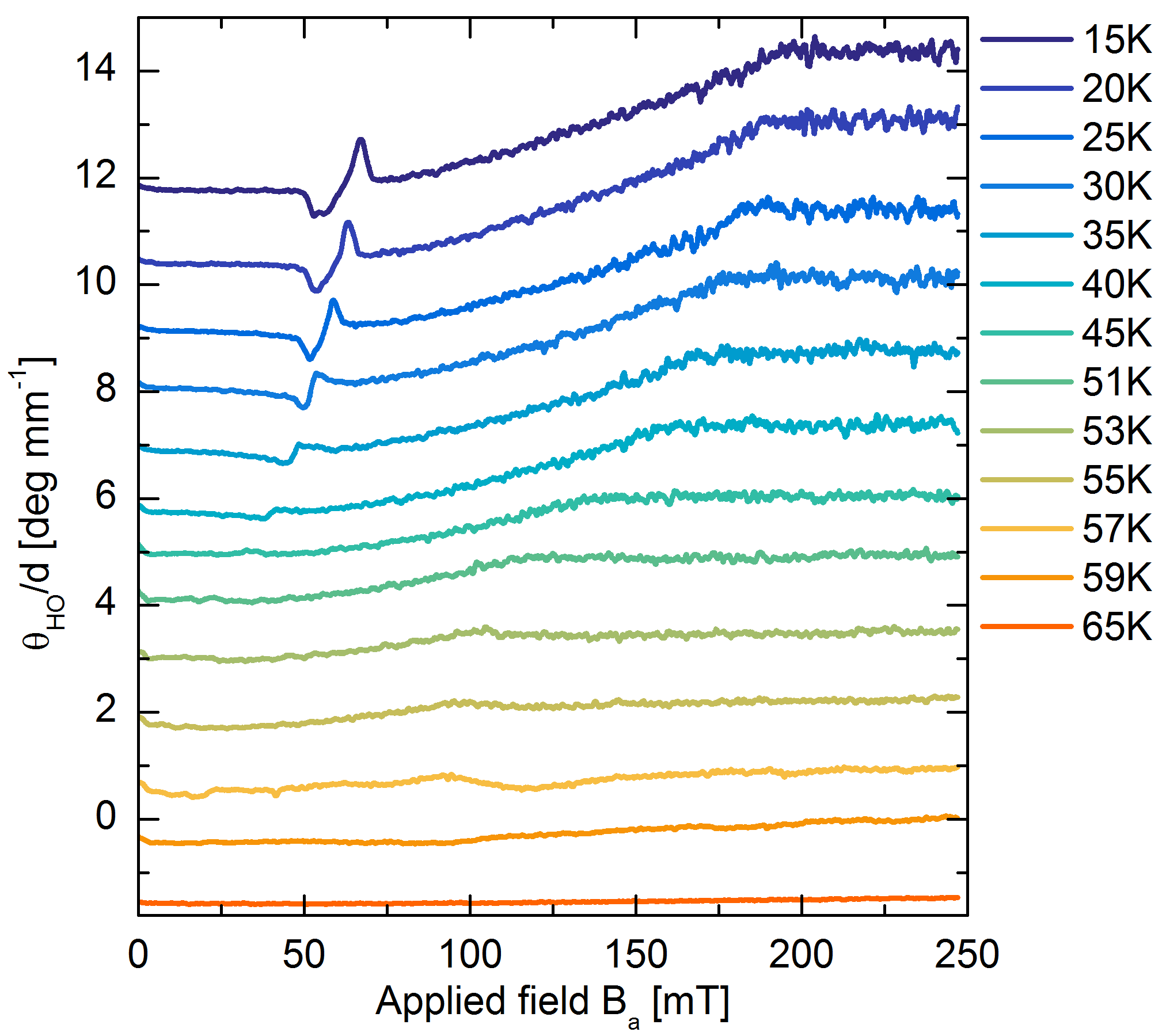} 
\caption{A field-even rotation 
$\theta_{\rm HO}$, shown here per sample thickness $d$,
is present due to the finite magneto-electric effect. The different curves are offset for clarity. The phase transition from conical to ferrimagnetic order again becomes apparent. The zig-zag feature at the helical-conical phase transition is due to hysteresis in $\theta_{\rm F}$.}
\label{fig:HOrotation}
\end{figure}

A higher-order magneto-optical rotation $\theta_{\rm HO}$ is anticipated owing to the magneto-electric nature of \COSO. Through the finite magneto-electric coupling, the natural optical activity is enhanced in the helimagnetic phase, as discussed above. With this coupling, a field-even optical rotation $\theta_{\rm HO}(T,(M^2(B_a^2))$ is expected, see Equ.\ \ref{eq:thetatot}.
We indeed find an appreciable polarization rotation symmetric in $B_a$, as shown in Fig.\ \ref{fig:HOrotation} 
At $15\,$K, the even rotation is about a factor $60$ smaller than the Faraday rotation. With increasing magnetization, the field-symmetric rotation $\theta_{\rm HO}$ increases until the ferrimagnetic phase is reached, where the rotation levels off. At the helical-conical phase transition a marked zig-zag feature is present. This feature is not a genuine higher-order rotation, but instead reflects the hysteresis in the Faraday rotation across the helical-conical first-order phase transition. 

\subsection{Magnitude of the magneto-optical effect}

At $15\,$K, we find a polarization rotation of around $165\,$deg/mm in the field-polarized 
ferrimagnetic phase above 195\,mT and a magneto-optical susceptibility of the order of $-10^4$\,rad/T$\cdot$m 
for smaller magnetic fields. 
In the following, we will show that both values are large but can be explained by the fact that the spin clusters are 
fully polarized already for moderate magnetic fields of 195\,mT.

For a field-independent magnetic susceptibility, the size of the magneto-optical response can be 
expressed by the Verdet coefficient $\mathcal{V}(\omega)$\,=\,$\beta(\omega)$\,$\cdot$\,$\chi_m$ (cf.\ Equ.\ \ref{eq:susceptibility}). 
In \COSO, it is thus possible to assign a Verdet coefficient of 
$-1.5\cdot 10^{4}\,$rad/T$\cdot$m to the helical phase at low temperatures and small fields, 
see Fig.\ \ref{fig:fluctuationsverdet}. 
Taking into account the demagnetizing field correction,\cite{criticalscaling,satoishii1989} 
we find $\mathcal{V}(540$\,nm)\,$\approx$\,-3$\cdot$10$^{4}\,$rad/T$\cdot$m, a value rivaling known strong magneto-optical rotators such as the paramagnets Tb$_3$Ga$_5$O$_{12}$ (with $\mathcal{V}$($1053\,$nm)\,$\approx$\,0.3$\cdot$10$^{4}\,$rad/T$\cdot$m 
at $4.2\,$K)\cite{yasuhara2007} and Cd$_{1-x}$Mn$_{x}$Te ($\mathcal{V} \lesssim 9\cdot$10$^{4}\,$rad/T$\cdot$m at $77\,$K depending on $\omega$ and stoichiometry).\cite{gaj1978} For these paramagnets, the large size of $\mathcal{V}$ solely originates in $\beta(\omega)$.\cite{pershan1967} In contrast, the large magneto-optical response of \COSO can be attributed to the large magnetic susceptibility
at low fields and the strong magnetization at larger fields.

The magnetization\cite{jwbos2008}, of \COSO shows that the $S$\,=\,1 clusters are fully aligned along 
the applied field in the ferrimagnetic phase, i.e., already for moderate magnetic fields of 0.2\,T.\@ 
This reflects that the magnetocrystalline anisotropy is weak in this cubic magnet with small spin-orbit coupling. As a result, the helical or conical domains can be easily reoriented,\cite{chizhikov2015} even when the magnetic field 
is applied along the $\langle$111$\rangle$ hard axis. This in turn is reflected in a large magnetic susceptibility $\chi_{m}$, and hence a large magneto-optical susceptibility $\chi_{\rm MO}$.

For the field-polarized ferrimagnetic phase, strong magneto-optical effects are naturally expected. 
In fact, roughly similar (volume) magnetizations and polarization rotations have been observed 
in ferromagnetic cuprates such as the 2D compounds K$_2$CuF$_4$ [\onlinecite{laiho1976}] and (CH$_3$NH$_3$)$_2$CuCl$_4$ [\onlinecite{arend1975}], showing rotations of $36\,$deg/mm and $50\,$deg/mm, respectively. 
This shows that the magneto-optical coupling $\beta(\omega)$ in \COSO is of equal order of magnitude as in these materials. 

The microscopic magneto-optical interaction $\beta(\omega)$ depends on the difference in dipole transition strength 
for the $\sigma_{+}$ and $\sigma_{-}$ transitions at $\omega$, 
where $\pm$ refers to right/left circularly polarized light, respectively.\cite{pershan1967} 
A \textit{larger} difference in transition strength leads to a \textit{stronger} rotation. 
In a ferromagnet with full spin polarization, only one of the transitions $\sigma_{\pm}$ is preferentially allowed 
due to angular momentum conservation. 

In contrast, a perfect antiferromagnet with zero magnetization shows equal transition strengths and 
thus no polarization rotation. 
In \COSO, one finds a ferromagnetic alignment of the $S$\,=\,1 clusters but antiferromagnetic alignment between 
Cu-I and Cu-II sites within a given cluster, see Fig.\ \ref{fig:splittings}. 
However, the excitation energies are different for Cu-I and Cu-II sites, which holds for both, the crystal-field excitations and charge-transfer excitations. In addition, the Cu-I and Cu-II sites are present in a ratio of 1:3. Accordingly, the cancellation typical for a simple antiferromagnet 
does not occur in \COSO. 
Moreover, the transition dipole strength of both types of excitations is relatively large, which magnifies the  magneto-optical response.  Whether the different optical transitions have the same rotational sense or not 
cannot be answered based on a single-wavelength measurement and remains open for further studies.

\section{Conclusions}

We have shown how a variety of linear optical properties can be used to probe inversion and time-reversal breaking properties of the chiral magnetic cuprate \COSO. The broken inversion symmetry at the two crystallographically inequivalent Cu sites leads to a finite dipole matrix element for local crystal-field excitations. These orbital excitations were observed below the charge-transfer gap, in the energy range from $1.0$ to $2.0$\,eV.\@ The transmission window found between 
the crystal-field and the charge-transfer absorption regions allowed us to measure the optical polarization
rotation across the various magnetic phase transitions in \COSO. The zero-field optical rotation, corresponding to the natural optical activity of the crystal, is enhanced upon entering the helimagnetically ordered phase. This observation provides evidence for a finite magneto-electric coupling in the helimagnetic phase of \COSO. 

In the presence of an external magnetic field, the finite magnetization leads to a sizable Faraday rotation. The ease of domain reorientation by the external magnetic field was discussed to be at the origin of the large magneto-optical susceptibility in the helical and conical phases. 
The large Faraday rotation observed in the field-polarized ferrimagnetic phase agrees with 
results obtained on ferromagnetic cuprates.
The Faraday rotation provides a sensitive tool to conveniently probe the nature of the various magnetic phase transitions in \COSO, including the subtle first-order nature of the helimagnetic-paramagnetic phase transition. From the magneto-optical data we reconstructed  the phase diagram of \COSO as a function of magnetic field and temperature, including the skyrmion lattice phase.

\section{acknowledgement}
The authors are grateful to M. Garst, A. Rosch, P. Becker, L. Bohat\'{y}, and T. Lorenz for insightful discussions. We also thank P. Padmanabhan for critical reading of the manuscript. Part of this work was financially supported by the Bonn-Cologne Graduate School of Physics and Astronomy (BCGS).

\section*{Appendix}
\subsection*{Temperature dependence of ellipsometry data}

\begin{figure}[b]
\center
\includegraphics[scale=1]{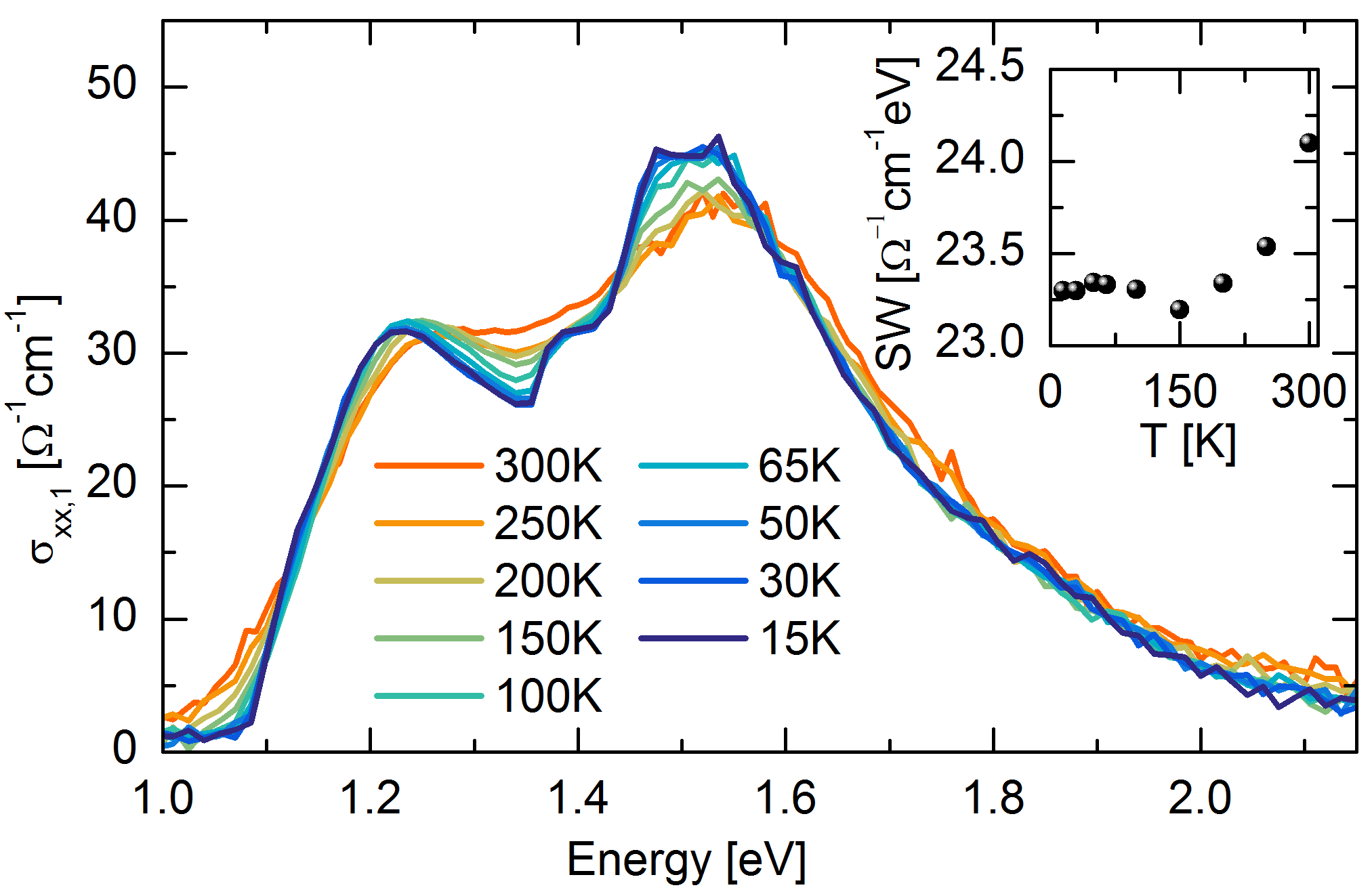}
\caption{Optical conductivity in the region of the crystal-field excitations for different temperatures. The inset depicts the spectral weight (integral of the optical conductivity between $1.0$ and $2.0\,$eV) as a function of temperature.}
\label{fig:COSO-ellip-Tdep}
\end{figure}

The temperature dependence of the optical conductivity in the region of the crystal-field excitations is shown in more detail in Fig.\ \ref{fig:COSO-ellip-Tdep}. The inset shows the temperature dependence of the spectral weight of this region, \ie the integral of $\sigma_1(\omega)$ from $1$ to $2$\,eV.\@ We find a slight increase of the spectral weight with increasing temperature. The onset of magnetic order at 58\,K has no significant effect on the spectral weight of the crystal-field excitations. 

\end{document}